\documentclass[twocolumn,showpacs,amsmath,amssymb,sort]{revtex4}

\usepackage{dcolumn}
\usepackage{bm}

\usepackage{amsmath}
\usepackage{amsfonts}
\usepackage{amssymb}
\usepackage{array}
\usepackage[dvips]{graphicx}

\usepackage{color}

\usepackage{ulem}
\usepackage{xcolor}
\definecolor{DarkGreen}{rgb}{0,0.7,0}
\definecolor{DarkRed}{rgb}{0.5,0.0}
\definecolor{DarkViolet}{rgb}{0.5,0.0,0.5}
\definecolor{highlighted}{rgb}{0.816,0.816,1.0}
\definecolor{redeb}{rgb}{1.,0.,0.}

\usepackage{epstopdf}

\newcommand{\EB}[1]{{\color{redeb} {#1}}} 
\renewcommand{\EB}[1]{#1} 

\newcommand{\Xout}[1]{}

\newcommand{\Ci}{\mathrm{i}} 	
\newcommand{\e}{\mathrm{e}} 	
\newcommand{\diff}{\mathrm{d}}
\newcommand{\pdiffer}[2]{\frac{\partial #1}{\partial #2}}

\newcommand{\bra}[1]{\langle #1 \vert}
\newcommand{\ket}[1]{\vert #1 \rangle}

\newcommand{\erwartungswert}[1]{\langle #1 \rangle}

\newcommand{\Order}{\mathcal{O}}

\newcommand{\op}{\hat}

\newcommand{\opH}{\op H}

\newcommand{\opc}{\op c}
\newcommand{\opcd}{\op c^\dagger}
\newcommand{\opcp}{\op c^{\phantom{\dagger}}}

\renewcommand{\Re}{{\mathrm{Re}}}

\renewcommand{\emph}[1]{\textit{#1}}

\newcommand{\R}[1]{\mathrm{#1}}
\newcommand{\SD}{{\textsc{sd}}}
\newcommand{\VSD}{{V_\SD}}

\newcommand{\Vg}{{V_\R{g}}}

\newlength{\graphiclength}
\graphicspath{.} 

\begin{document}

\author{A. Bransch{\"a}del}
\affiliation{Institut f\"ur Theorie der Kondensierten Materie, Karlsruhe Institute of Technology, 76021 Karlsruhe, Germany}
\author{E. Boulat}
\affiliation{Laboratoire MPQ, CNRS UMR 7162, Universit\'e Paris Diderot, 75205 Paris Cedex 13}
\author{H. Saleur}
\affiliation{Institut de Physique Th\'eorique, CEA, and CNRS, URA2306, Gif Sur Yvette, F-91191}
\affiliation{Department of Physics, University of Southern California, Los Angeles, CA 90089-0484}
\author{P. Schmitteckert}
\affiliation{Institute of Nanotechnology, Karlsruhe Institute of Technology, 76344 Eggenstein-Leopoldshafen, Germany }

\begin{abstract}
We present a method to determine the  shot noise in quantum systems from knowledge of their  time evolution - the latter being obtained 
using numerical simulation techniques. While our ultimate goal is the study of interacting systems, 
the main issues for the numerical determination of the noise do not depend on the interactions. 
To discuss them, we concentrate on  the single resonant level model,  
which consists in a single impurity attached to non-interacting leads, with spinless fermions. 
We use exact diagonalisations (ED) to obtain time evolution, and are able to use known  analytic results as benchmarks. 
We obtain a complete characterization 
of  finite size effects at zero frequency, where we  find that the finite size corrections  
scale $\propto G^2$, $G$ the differential conductance. We also discuss finite frequency noise, 
as well as the effects of damping in the leads. 
\end{abstract}

\title{Numerical Evaluation of Shot Noise using Real Time Simulations}
\maketitle



The study of current fluctuations in nanodevices such as quantum point contacts and tunnel junctions is deeply 
connected with some of the most important physical questions. These include the  nature of fundamental excitations 
in strongly interacting electronic systems \cite{Reznikov95,Kumar96}, the possibility of fluctuation theorems out 
of equilibrium \cite{Esposito08}, and the time evolution of many body entanglement \cite{Klich09a,Klich09b}. 
Experimental progress in this area has been swift - second and third cumulants have been measured in several systems \cite{Reulet03,Bomze05}, 
shot noise of single hydrogen molecules have been measured \cite{Djukic_vanRuitenbeek:NL2006},
and even the full counting statistics has been obtained in semi conductor quantum dots \cite{Gustavsson06}. 

On the theory side, the free case  has given rise to a lot of analytical studies 
\cite{Levitov93,Levitov96,Klich09a,Klich09b}, but progress on the most interesting situations - far from 
equilibrium and with strong interactions - has been difficult (see \cite{Gogolin08} for a review).  Over the years,   
extensions of the Bethe ansatz to study transport properties  have been proposed \cite{Fendley96, Konik01, Mehta08,Chao10}, which 
might open the road to important progress.

The  numerical situation is somewhat similar. $I-V$ characteristics had remained inaccessible  until very recently, 
where  methods based on time dependent DMRG   have been successfully used to tackle some simple nanostructures \cite{Boulat08}. 

It seems reasonable to expect that time dependent DMRG results could also be used to determine current fluctuations, 
which could also, in some setups, be determined analytically \cite{Weiss01}. In order to reach this ambitious 
goal, it is crucial to be able to extract cumulants - in particular the shot noise - from real time simulation 
methods. We propose in this paper  a  method to do this. 

The main problem in the determination of the noise - and the main emphasis  of our work - is  the finite size 
analysis of the results of non-equibrium correlation functions for  finite systems. To concentrate on  this aspect,  
we only discuss  results for the non-interacting resonant level model (RLM) 
where the numerical data can be obtained using exact diagonalisation (ED) techniques. Since in this specific case 
there are straightforward analytical solutions of the problem, we can check in great detail the reliability of our 
approach.
We want to emphasize however  that the method to be described below  is independent of the underlying numerical algorithm, 
and can therefore also be used in combination with, 
e.g., time dependent Density Matrix Renormalisation Group (td-DMRG). We will report on this in the case of the interacting 
resonant level model in a forthcoming paper.

\begin{figure}[b]
\begin{center}
   \includegraphics[width=\graphiclength]{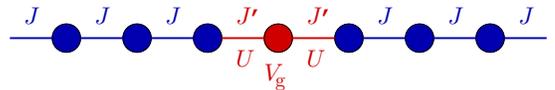}
   \caption{Sketch of the IRLM model. $J'$ denotes the coupling of the impurity to the left and right lead, 
   $J$ is the hopping element in the leads, $U$ is the interaction on the contact link, and $V_{\mathrm g}$ is a gate voltage.  
   In this work we set $V_{\mathrm g}=0$ (the dot level is on resonance), and $U=0$ (non-interacting case, which allows to treat the problem via ED).} 
\label{fig:IRLM}
\end{center}
\end{figure}

We note that prior to our work, a numerical study of the full counting statistics for another non interacting model 
appeared in \cite{Schoenhammer07}. The method used there is however entirely specific to the free case, and uses 
intermediate analytical results  from \cite{Levitov93}. Our approach, in contrast, is based directly on the `experimentally' measured  time evolution of 
the current.

To make things concrete, we start by giving the Hamiltonian of our test  system. It  is composed of a structure 
($\op H_\R S$) attached to leads ($\op H_\R L$) described in  real space by
\begin{flalign}
\label{eq:Hamiltonian1}
   \op H &= \op H_\R L + \op H_\R S, \\
\label{eq:Hamiltonian2}
   \op H_\R L &= 
       -J \sum_{m=-\infty}^{-2} \opcd_{m+1}\opcp_m
       -J \sum_{m=1}^{\infty} \opcd_{m+1}\opcp_m+\mathrm{h.c.}, \\
   \op H_\R S &= -J'\sum_{m=\pm 1}(\opcd_{m}\opcp_0+\opcd_{0}\opcp_m) + V_\R g \op n_0 \nonumber \\
       & \phantom{=} +U\sum_{m=\pm 1} \Big(\op n_m-\frac 1 2\Big)\Big(\op n_0 -\frac 1 2\Big),~~ \op n=\opcd\opc,
\label{eq:Hamiltonian3}
\end{flalign} 
cf. Fig.~\ref{fig:IRLM}.
The nearest-neighbour hopping matrix elements in the leads and the coupling of the structure to the leads are given by 
$J$ and $J'$. 
In the remainder of this work we concentrate on the resonant case at zero gate voltage $V_{\R g}=0$ and half filling. 
Since we want to compare the numerical data with analytical results, we furthermore restrict ourselves to the non-interacting 
case with $U=0$.

To prepare the system in a state with finite current through the structure, we add a charge imbalance operator 
$\op Q = \VSD \left( \op N_{\R L} - \op N_{\R R} \right) / 2$ to the Hamiltonian and calculate the initial state as the ground 
state $\ket{\Psi(t=0)}=\ket{\Psi_0}$ of $\opH+ \op Q$. Here, $\op N_\R L$ ($\op N_\R R$) counts the particle number in the left 
(right) lead.
We then perform the time evolution with
the Hamiltonian without $\op Q$.
The time evolution is performed by means of the time evolution operator $\op U=\e^{-\Ci \op H t}$, 
while all expectation values are evaluated with respect to the initial state $\ket{\Psi_0}$.  For details see
\cite{SchmitteckertSchneider:HPC06,SchneiderSchmitteckert06,BoulatSaleurSchmitteckert2008,AlHassaniehFeiguinRieraBusserDagotto06,SilvaHeidrichMeisnerFeiguinEtAl08,KirinoFujiiZhaoUeda08,HeidrichMeisnerFeiguinDagotto09,BranschadelSchmitteckert09}.

\begin{figure}[t]
 \graphicspath{{./fig/}}
 \begin{footnotesize}\input{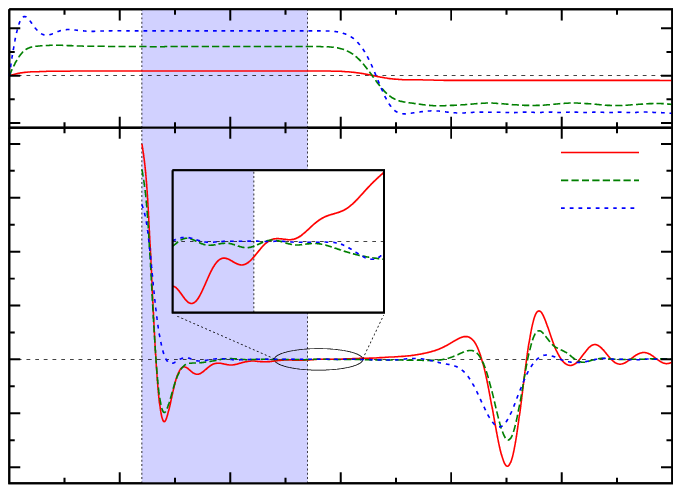}\end{footnotesize}
 \caption{Time dependent current $I(t)$, and current correlation function $S(t,t_{\text{min}})$ with $t_{\text{min}}=12$, in the 
non-interacting resonant level model RLM, with tight-binding leads and a finite system size of $M=60$ lattice sites, for different 
values of the bias voltage $\VSD$. The $I(t)$ curves show the three time regimes given by the settling time $t_\R S$ and the recurrence 
time $t_\R R$. The \colorbox{highlighted}{highlighted} time domain indicates the integration range $[t_\text{min},t_\text{max}]$. The 
inset demonstrates an additional subtility: the correlation function shows finite size reflection effects on the time scale 
$t-t_\R{min}\gtrsim t_\R R /2$, which imposes an additional restriction on $t_{\R{max}}$.}
\label{fig:timeranges}
\end{figure}

The current operator $\op I_{m}$ for the current at bond $m$ is given by
\begin{equation}
 \op I_{m}(t)=\Ci\frac e\hbar J_m 
              \left[\opcd_j(t)\opcp_{j+1}(t) -\opcd_{j+1}(t)\opcp_j(t)\right].
\end{equation} 
We define the current operator as an average over the current on the left and the right contact of the nanostructure
\begin{equation}
 \op I(t)=\frac 1 2 \left[ \op I_{-1}(t)+ \op I_0(t) \right].
\end{equation} 
The expectation value of $\op I(t)$ in the RLM for $J'=0.4J$ and for some values of $\VSD$ is shown in the upper part of Fig.~\ref{fig:timeranges}. 
Effects like the finite settling time $t_\R S$ and the finite transit time $t_\R R$ as well as the $I$-$\VSD$-characteristics 
have been discussed before in great detail \cite{SchmitteckertSchneider:HPC06,wingreen93,BranschadelSchmitteckert09}.

\setlength{\fboxsep}{1pt}%

Shot noise is defined as the zero-temperature contribution to noise in a transport state. To obtain the noise power spectrum from a 
real time simulation, the current-current correlations in the time domain 
\begin{align}
 S(t,t') &= \frac 1 2 \erwartungswert{\Delta \op I(t) \Delta \op I(t') +
                        \Delta \op I(t') \Delta \op I(t)} \\
         &= \Re\erwartungswert{\Delta \op I(t) \Delta \op I(t')}
\end{align} 
have to be calculated in a non-equilibrium zero-temperature state, where $\Delta \op I(t)=\op I(t)-\erwartungswert{\op I(t)}$ 
\cite{LandauLifshitz, Imry1997}. Therefore, the time dependent expectation value
\begin{equation}
	\erwartungswert{\Delta \op I(t) \Delta \op I(t')} = \bra{\Psi_0}\e^{\Ci\opH t}\Delta\op I\e^{-\Ci\opH(t-t')}\Delta\op I \e^{-\Ci\opH t'}\ket{\Psi_0}
\end{equation} 
has to be evaluated. In a steady state the correlation function must fulfil $S(t,t')\equiv S(t-t')$. Then the noise power can be defined 
as the Fourier transform
\begin{equation}
 2\pi\delta(\omega+\omega')S(\omega) = \erwartungswert{
          \Delta \op I(\omega) \Delta \op I(\omega') 
        + \Delta \op I(\omega') \Delta \op I(\omega)},
\end{equation} 
where
\begin{align}
 S(\omega) &= 2\int\limits_{-\infty}^\infty \diff t\, \mathrm e^{\mathrm i \omega t} S(t,t'=0) \\
           &= 4\Re \int\limits_{0}^\infty \diff t\, \mathrm e^{\mathrm i \omega t} S(t,t'=0).
\end{align} 
The right-hand side of the equation accounts for the symmetry $S(t-t')=S(t'-t)$. In a steady state, of course, this expression should be 
independent of the choice of the time $t'$
\begin{equation}
 S = 4 \Re\int\limits_{t'}^\infty \diff t\, \mathrm e^{\mathrm i \omega (t-t')} S(t,t')~~~\forall\; t'.
\label{eq:S}
\end{equation} 
In the zero-frequency limit $\omega=0$ this expression simplifies to
\begin{align}
 S\equiv S(\omega=0) &= 4 \int\limits_{t'}^\infty \diff t\, S(t,t') \\
             &= 4 \int\limits_{t'}^\infty \diff t\, \Re\erwartungswert{\Delta \op I(t) \Delta \op I(t')}.
\label{eq:S0}
\end{align} 
We now want to see whether the noise can be reliably obtained  in a real time simulation based on this formula. 

There are of course many obstacles. The first comes from the 
 calculation of a  nonequilibrium correlation function in the time domain from a real time simulation.
Because we are restricted to a finite system with $M$ lattice sites and hard walls,
 a steady transport state is not well defined. 
Instead, we make the attempt to calculate the time evolution from the initial non-equilibrium state $\ket{\Psi(t=0)}$ as described before
and look for a quasi-stationary time regime.
The ``switching'' of a finite source-drain voltage $\VSD$ at initial time causes a ringing of the current \cite{wingreen93}, cf. also Fig.~\ref{fig:timeranges},  
which decays exponentially within a settling time $t_\R S$. The current finally enters a plateau regime, 
where the size of the plateau is given by the recurrence time $t_\R R$ which is finite due to the finite size of the system 
\cite{SchneiderSchmitteckert06}, Fig.~\ref{fig:timeranges}. 

\begin{figure}[t]
 \graphicspath{{./fig/}}
 \begin{footnotesize}\input{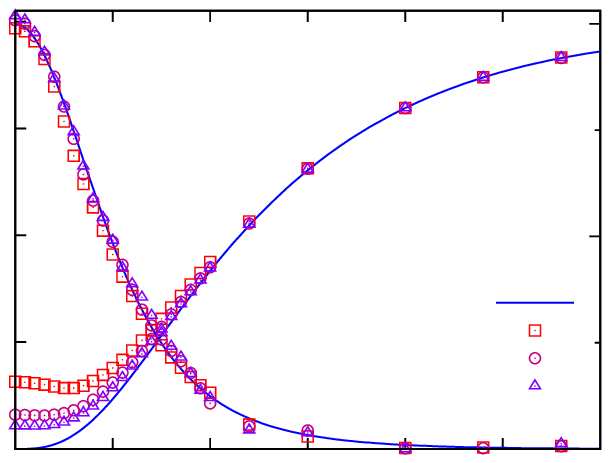}\end{footnotesize}
 \caption{Noise $S$ and squared differential conductance $G^2$ for $J'=0.4J$. The blue lines represent the analytic values 
   obtained using the Landauer--B\"uttiker approach. 
   The finite size of the system introduces an additional noise proportional to $G^2/M$ (for discussion see text).
 }
\label{fig:systemsizes}
\end{figure}

Having obtained results for the system   ``close'' to a steady state, to obtain the quantity $S$ we now have  to evaluate the 
integral \eqref{eq:S0} in a limited time range
\begin{equation}
 S_{\R{num}} =
   4 \int\limits_{t_\R {min}}^{t_\R {max}} \diff t\, \Re\erwartungswert{\Delta \op I(t) \Delta \op I(t_\R {min})}_\Psi
\label{eq:S_num}
\end{equation} 
where $t_\R{min}> t_\R S$ and $t_\R{max}< t_\R R$. In a hypothetical situation with a system of infinite size where 
$t_\R R \rightarrow \infty$ the contribution of 
$\int_{t_\R {max}}^\infty \diff t\, \Re\erwartungswert{\Delta \op I(t) \Delta \op I(t_\R {min})}_\Psi$ can be neglected if 
$\Re\erwartungswert{\Delta \op I(t) \Delta \op I(t_\R {min})}_\Psi$ is small for $t>t_\R{max}$ as compared to the mean value in the range 
$t_\R{min}<t<t_\R{max}$. One therefore has to choose the size of the system big enough to ensure the correlation function to drop to zero 
within the recurrence time. 

The finite recurrence time $t_\R R$ introduces a finite cutoff frequency
$\omega_\text{cut}=2\pi/{t_\R R}\propto 1/M$. This is the main problem we encounter. In contrast to the situation of
infinite leads, where zero frequency noise vanishes  without applied voltage, we
now find a contribution to the zero voltage shot noise of the order of 
$S( \omega_\text{cut} )$! The low frequency domain is the most interesting for the kind of problems we wish to study: low frequency is low 
energy and thus strong coupling  \EB{between impurity and leads}.
\begin{figure}[t]
 \graphicspath{{./fig/}}
 \begin{footnotesize}\input{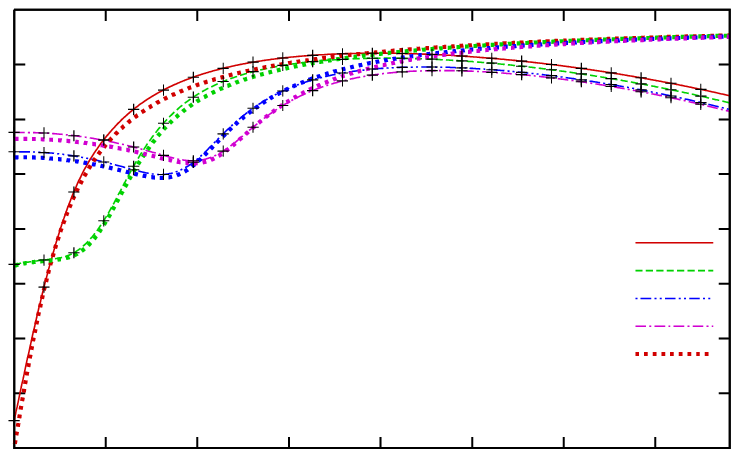}\end{footnotesize}
 \caption{Noise $S(\omega)$ vs.\ frequency $\omega$, both rescaled with respect to the width of the conductance peak $\Gamma=(2J'/J)^2$, for 
 different values of the bias voltage. The lines going through the numerical values (represented by crosses) are just guides for  the eye. The dotted 
 lines correspond to the analytical result for the wide band limit.
 }
\label{fig:SomegaAnalytical}
\end{figure}

The magnitude of the finite size effects for the type of systems that can be studied today is far from negligible. 
On fig.~\ref{fig:systemsizes} we give results for  the shot noise  $S_\text{num}$ obtained  for different system sizes of $M$ lattice sites,
as well as the expected result  $S$ in  the thermodynamic limit obtained from the Landauer--B\"uttiker approach (this is discussed in more details in 
the appendix). \xout{Obviously,} While the results measured for finite size and the asymptotic results agree at large voltages, 
there is a marked difference at small voltages, with an offset at vanishing $V_\text{SD}$. 
On the figure, we also represented  the finite size correction
\begin{equation}
 \Delta S_\text{num} = S_\text{num}-S
\end{equation} 
rescaled by the system size $M$. 
For different values of $M$ the rescaled finite size corrections $M\times \Delta S_\text{num}$ collapse very well on a single curve, 
indicating that the main  finite size effects scale linearly with $1/M$ in the considered parameter regime. 
One may expect that the cut off given by the finite size of the leads corresponds to an effective finite temperature $\sim M^{-1}$
resulting in a low voltage offset $\sim G/M$.
However, we find $\Delta S_\text{num} \propto G^2$ with the differential conductance $G(\VSD)=\diff I(\VSD)/\diff\VSD$. 


\bigskip

To understand the behavior of  $\Delta S_\text{num}$, we consider  the full frequency dependence of the shot noise. 
It can easily be obtained analytically  in the wide band limit  -  see  the appendix, Eq.~\eqref{eq:WBL}. 
For values of $J'/J \ll 1$, the numerical results obtained for the model with cosine dispersion relation should be consistent with the analytical 
result as long as the considered frequency is small  compared to the band width. This is illustrated on fig.~\ref{fig:SomegaAnalytical}. 
There, the frequency dependent noise is obtained via
\begin{equation}
 S_\text{num}(\omega) = 4 \Re\int\limits_{t_\text{min}}^{t_\text{max}} \diff t\, \mathrm e^{\mathrm i \omega (t-t_\text{min})} S(t,t_\text{min})
\label{eq:Snum_omega}
\end{equation} 
for different values of the bias voltage $\VSD$. For big values of $\omega$, the  effects of the band curvature are quite marked - as can be seen 
by the departure of the various guide lines from the dotted lines representing the analytic wide band limit results . 
\begin{figure}[t]
 \graphicspath{{./fig/}}
 \begin{footnotesize}\input{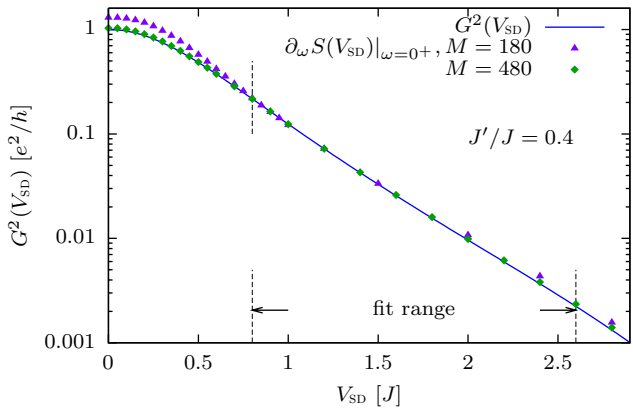}\end{footnotesize}
\caption{Slope of the frequency dependent shot noise in the limit $\omega\rightarrow 0^+$, rescaled to fit with $G^2$, for different system sizes. 
In the low voltage regime we find finite size effects.
}
\label{fig:Slope}
\end{figure}

To understand the voltage dependency of the finite size corrections, we consider  the low frequency behaviour  of the analytical results in the 
wide band limit where we find
\begin{equation}
 S(\omega>0,\VSD)= S(0,\VSD)+\Delta S(\omega,\VSD)+\Order(\omega^2)
\end{equation} 
with the correction in first order with respect to $\omega$
\begin{equation}
 \Delta S(\omega,\VSD) \propto G^2(\VSD) \omega.
\end{equation} 
For the system with finite band width, we have checked this expression by extracting the slope $\partial S(\omega,\VSD)/\partial \omega$ in the 
limit $\omega\rightarrow 0^+$ from the numerical data. Again we find good agreement with $G^2$ in a voltage regime where finite size effects can 
be neglected, Fig.~\ref{fig:Slope}.

%

Inserting the cutoff frequency now leads to the expression
\begin{equation}
 \Delta S(\omega_\text{cut},\VSD)\propto \frac 1 M G^2(\VSD)
\label{eq:finite_size_S}
\end{equation} 
which is in good agreement with $\Delta S_\text{num}(\VSD)$, cf. Fig.~\ref{fig:systemsizes}. 

\begin{figure}[t]
 \graphicspath{{./fig/}}
 \begin{footnotesize}\input{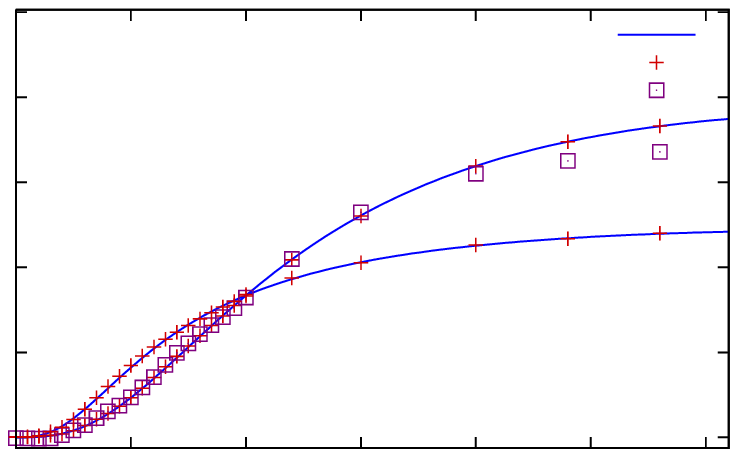}\end{footnotesize}
 \caption{Shot Noise $S$  as function of the source drain voltage $\VSD$ in the non-interacting resonant level model. 
 The analytical result was obtained using the Landauer--B\"uttiker theory, Eq.  \eqref{eq:noise_LB}, 
 while the numerical result is computed for systems of different finite sizes of  
 $M=120...180$ lattice sites with a subsequent linear extrapolation of  $1/M\rightarrow 0$. 
 The two curves correspond to different couplings  $J'$ of the impurity to the leads. 
 Furthermore, we used DBC in order to effectively increase the system size. 
 Here, the system size was fixed to $M=60$ lattice sites. 
 For weak damping and for not too big values of $\VSD$, 
 we find very good agreement with the undamped case.
 }
\label{fig:analytical_vs_exactDiagonalisation}
\end{figure}

Using our knowledge of the finite size correction, we can now control the extrapolation of numerical data:
in Fig.~\ref{fig:analytical_vs_exactDiagonalisation} we show the results obtained using
linear extrapolation $1/M\rightarrow0$ for $J'=0.3J$ and $J'=0.4J$. We find indeed 
very good agreement with the analytical result.


The non-interacting case is of course very simple to calculate numerically
(regardless of the possibility of the Landauer--B\"uttiker treatment). The numerical
main effort consists in the exact diagonalisation of the $M\times M$ Hamiltonian
matrices as well as the calculation of the time evolution which involves
the multiplication of  $M\times M$  matrices.
Including interaction spoils this approach. Instead, one has to resort to
approximative time evolution schemes using methods for correlated electrons. 
In \cite{BranschadelSchmitteckert09} we showed that using
Wilson leads, or damped boundary conditions (DBC), respectively, with a weak damping
constant allows one to effectively increase the system size to $M_\text{eff} > M$
lattice sites without changing $M$, where a rough estimate for $M_\text{eff}$ has
been given as a function of the damping constant $\Lambda$ and the length of the
damped leads $M_\Lambda$
\begin{figure}[t]
 \graphicspath{{./fig/}}
 \begin{footnotesize}\input{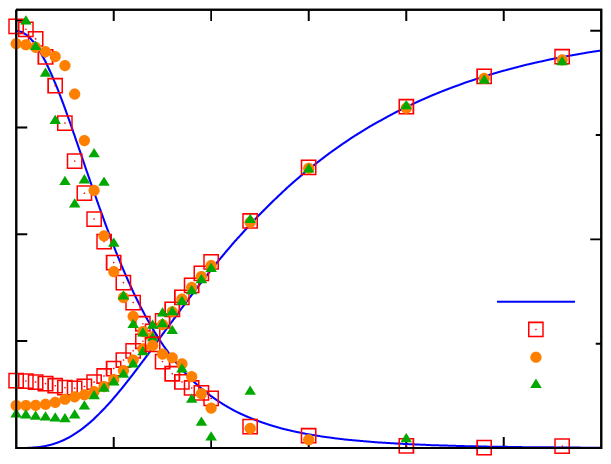}\end{footnotesize}
 \caption{
   Noise $S$ and squared differential conductance $G^2$ for $J'=0.4J$. The blue lines 
   represent the analytic values obtained using the Landauer--B\"uttiker approach. 
   The system size is fixed to $M=60$ lattice sites, while at the boundaries, the hopping matrix elements are exponentially damped with 
the damping constant $\Lambda^{-1/2}$ on $M_\Lambda$ links. This results in an effectively enlarged system with $M_\text{eff}$ lattice sites. 
The finite size correction $\Delta S_\text{num}$, here rescaled by $M_\text{eff}$, again collapses on a single curve for different $M_\text{eff}$, 
and is still proportional to $G^2$.
 }
\label{fig:Meff}
\end{figure}
\begin{equation}
 M_\text{eff}\approx M-2M_\Lambda+\frac 4{\ln \Lambda}\big(\Lambda^{M_\Lambda/2}-1\big).
\label{eq:Meff}
\end{equation}
We now use this estimate to perform the linear
extrapolation to infinite system size, where we additionally adjust the estimate by
fixing the extrapolated value to analytic results (cf., e.g., \cite{BlanterButtiker99})
\begin{equation}
 S(\VSD=0) = 0.
\end{equation} 
To verify this approach we performed calculations for a non-interacting system with
$M=60$ lattice sites and DBC, for $J'=0.4J$. For the damped leads we used different
combinations of $\Lambda$ and $M_\Lambda$, where we used values for the damping
constant in the range $\Lambda^{-1/2}\in [0.93, 1.0]$ for damped leads of
$M_\Lambda=0\ldots26$ lattice sites (while keeping the total number of lattice sites
$M$ fixed!). The estimate for the effective system size, Eq.~\eqref{eq:Meff}, is checked by looking at the scaling behaviour of the finite size 
correction $\Delta S_\text{num}$, where we now find linear scaling $\propto 1/M_\text{eff}$, cf. Fig.~\ref{fig:Meff}. 

The result is shown in
Fig.~\ref{fig:analytical_vs_exactDiagonalisation}. We find remarkably good agreement
with the analytical result, while we have to point out that, for values of the bias
voltage in the order of the band width, the approach fails, which has to be expected
since the estimate of the system size only works in a limited voltage range
\cite{BranschadelSchmitteckert09}. Additionally we find the numerical data to be very noisy depending on the respective configuration of the damping conditions.

To conclude, we have introduced a new way of extracting the finite bias shot noise from real time evolution calculations. 
Very accurate quantitative results are possible as long as finite size effects are treated properly. 
The possibility of effectively enlarging the system size using Wilson leads has been successfully tested for a limited voltage range. 
Our results for shot noise in the RLM show very good agreement with results obtained from the Landauer--B\"uttiker approach
for zero as well as finite frequencies.
Additionally we confirm a $G^2$ dependent scaling of the finite size error in the zero frequnecy regime.
  
The concept is not restricted to non-interacting fermions and could be implemented using 
numerical methods for interacting quantum systems.
In this case as well, we expect the finite size corrections to go as $1/M$ because the cutoff frequency $\omega_\text{cut}$ has the same dependence. 
We note however that the prefactor might not be $G^2(V_\text{SD})$ exactly: the jury is still out on the small frequency behavior of the noise in the presence of interactions.\cite{Lesage97,Chamon95}

We acknowledge the support by the DFG Center  
for Functional Nanostructures (CFN), project B2.10.
\appendix

\section{Analytical results}

For the non-interacting system in the thermodynamic limit, Eqns.~(\ref{eq:Hamiltonian1}-\ref{eq:Hamiltonian3}), with $U=0$, one can derive analytic 
results for the differential conductance and the shot noise from single particle scattering states. The energy dependent single particle tunneling 
probability $\tau$ is given as
\begin{equation}
	\tau(\epsilon)=\frac{1-\epsilon^2/(4J^{\EB{2}})}{1+(\epsilon-\Vg)\Big[(J^2-\EB{2}J'^2)\epsilon-J^2 \Vg\Big]/(4J'^{\EB{4}})}.
\end{equation} 
Using the Landauer approach (cf., e.g., \cite{BlanterButtiker99}) one obtains for the dc current $I$ and the noise $S$ at zero frequency and for a 
symmetric voltage drop $\VSD$
\begin{align}
	I(\VSD) &= \int_{-\VSD/2}^{\VSD/2}\diff\epsilon\, \tau(\epsilon), \\
	S(\VSD) &= \int_{-\VSD/2}^{\VSD/2}\diff\epsilon\, \tau(\epsilon)(1-\tau(\epsilon)).
\end{align}
For the resonant case with $\Vg=0$ this results in
\begin{equation}
	G(\VSD) = \pdiffer{I}{\VSD} = \frac{1-\VSD^2/(16J^{\EB{2}})}{1+\VSD^2(J^2-\EB{
	2}J'^2)/(16J'^{\EB{4}})}, \\
\end{equation} 
\begin{widetext}
\newcommand{\TB}{T_\textsc{b}}
\begin{align}
S(\VSD)=\frac{1}{2}\bigg(1+\left(\frac{\TB}{4J}\right)^2\bigg)\Bigg[
\TB\bigg(1+3\left(\frac{\TB}{4J}\right)^2\bigg)\;\arctan\left(\frac{V}{\TB}\right)
-V\;\frac{1+3\left(\frac{\TB}{4J}\right)^2+2\left(\frac{V}{4J}\right)^2}{1+\left(\frac{V}{\TB}\right)^2}
\Bigg]
\label{eq:noise_LB}
\end{align}
where $\TB={4J'^2}/{\sqrt{J^2-2J'^2}}$ is a scale.
For the frequency dependent noise, we will contend ourselves by giving the result for the wide band limit only:
\begin{align}
	S(\omega,\VSD,\TB) =& \frac\TB 4 \Theta(\VSD-\vert\omega\vert) 
		\Big\lbrace 
			\Big[ \arctan\Big(\frac\VSD\TB\Big) + 
					\arctan\Big(\frac{\VSD-2\vert\omega\vert}\TB\Big) \Big ]
			+ \frac\TB{2\omega}\ln\Big(\frac{\TB^2+(\VSD-2\vert\omega\vert)^2}{\TB^2+\VSD^2}\Big)
		\Big\rbrace \nonumber \\
		&+\frac\TB 4 
			\Big\lbrace \arctan\Big(\frac{\VSD+2\vert\omega\vert}\TB\Big)
						  -\arctan\Big(\frac{\VSD-2\vert\omega\vert}\TB\Big)
			\Big\rbrace,
\label{eq:WBL}
\end{align}
where in this limit $\TB={4J'^2\over J}$. 

\end{widetext}


\end{document}